\documentclass{emulateapj}

\slugcomment{Accepted for publication in ApJ Letters}

\shorttitle{Coldest Imaged Companion of a Sun-Like Star}
\shortauthors{Thalmann et al.}

\begin{document}

\title{Discovery of the Coldest Imaged Companion of a Sun-Like
Star\altaffilmark{$\star$}}

\author{C. Thalmann\altaffilmark{1,2}, J. Carson\altaffilmark{1,3}, 
	M. Janson\altaffilmark{4}, M. Goto\altaffilmark{1}, 
	M. McElwain\altaffilmark{5}, S. Egner\altaffilmark{6},
	M. Feldt\altaffilmark{1}, J. Hashimoto\altaffilmark{7}, 
	Y. Hayano\altaffilmark{6}, T. Henning\altaffilmark{1},
	K.~W. Hodapp\altaffilmark{8}, R. Kandori\altaffilmark{7},
	H. Klahr\altaffilmark{1}, 
	T. Kudo\altaffilmark{7}, N. Kusakabe\altaffilmark{7},
	C. Mordasini\altaffilmark{1}, 
	J.-I. Morino\altaffilmark{7}, H. Suto\altaffilmark{7},
	R. Suzuki\altaffilmark{6}, M. Tamura\altaffilmark{7}}

\altaffiltext{$\star$}{Based on data collected at Subaru Telescope, which
	is operated by the National Astronomical Observatory of Japan.}
\altaffiltext{1}{Max Planck Institute for Astronomy, Heidelberg, Germany}
\altaffiltext{2}{E-mail: \texttt{thalmann@mpia.de}}
\altaffiltext{3}{College of Charleston, Charleston, South Carolina, USA.}
\altaffiltext{4}{University of Toronto, Toronto, Canada}
\altaffiltext{5}{Princeton University, Princeton, New Jersey, USA}
\altaffiltext{6}{Subaru Telescope, Hilo, Hawai`i, USA}
\altaffiltext{7}{National Astronomical Observatory of Japan, Tokyo, Japan}
\altaffiltext{8}{Institute for Astronomy, University of Hawai`i, Hilo, Hawai`i, USA}

\begin{abstract}
We present the discovery of a brown dwarf or possible
planet at a projected separation of 1.9$^{\prime\prime} =$
29\,AU around the star GJ~758, placing it between the separations
at which substellar companions are expected to form by core 
accretion ($\sim$5\,AU) or direct gravitational 
collapse (typically $\gtrsim$100\,AU).
The object was detected by direct imaging
of its thermal glow with Subaru/HiCIAO.
At 10--40 times the mass of Jupiter and a 
temperature of 550--640\,K, GJ~758~B constitutes one of the
few known T-type companions, and the coldest ever to be imaged in 
thermal light around a sun-like star.  Its orbit is likely eccentric
and of a size comparable to Pluto's orbit, possibly as a result of 
gravitational scattering or outward migration.  A candidate second
companion is detected at 1.2$^{\prime\prime}$ at one epoch.
\end{abstract}





\section{Introduction}

While the extrasolar planets and brown dwarf companions currently known
from direct imaging mark significant discoveries, it is important to
note that none of these systems present scenarios analogous to a 
Solar-like
system.  Almost all imaged substellar companions
either feature extreme orbital separations \citep[hundreds of AU,
e.g.][]{luhman2007,kalas2008}, star-like rather than planet-like 
temperatures
\citep[e.g.][]{marois2008}, or host stars at the extreme ends of the
mass spectrum \citep[A- and late M-type, e.g.][]{kalas2008,chauvin2005}.
The closest match is probably Gl~229~B, a brown dwarf orbiting an M1 
star at 40\,AU \citep{nakajima1995}.
This shows how a remarkable gap still exists between recent direct-imaging discoveries and the exploration of Sun-like systems.  

In this work, we present the discovery of a substellar companion with a mass
of 10--40\,$M_\mathrm{Jup}$, a temperature of 550--640\,K, and a projected
separation of 1.9$^{\prime\prime} =$ 29\,AU.
The estimated semi-major axis range, overlapping with our own Solar System's outer planet orbits, along with the Sun-like parent star, make it a superior laboratory for advancing our conventional wisdom of planet or brown dwarf formation among solar-like systems.

\section{Observations and data reduction}

\begin{figure}
\centering
\includegraphics[width=\linewidth]{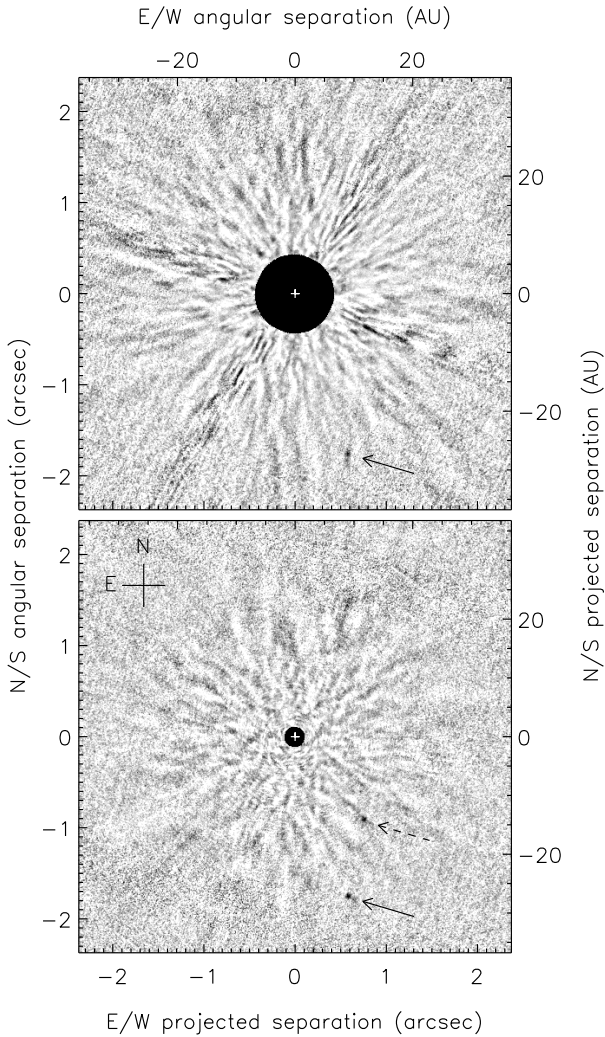}
\caption{Signal-to-noise maps of the discovery images for 
  GJ~758~B after ADI data reduction, for the May 2009 
  (\textbf{top}) and August 2009  (\textbf{bottom}) observations. 
  The greyscale stretch is $[-1\,\sigma,5\,\sigma]$.  The S/N ratio
  is calculated in concentric annuli around the star. 
  GJ~758~B is marked with solid
  arrows.  A possible second object in the August images is pointed
  out with dashed arrows.  The white plus sign marks the location
  of the host star GJ~758; the black disks designate the regions in
  which the field rotation arc is insufficient for ADI.}
\label{f:LOCI}
\end{figure}

GJ~758 is a G9-type star, located at a distance of 15.5\,pc \citep{gray2003,HIPPARCOS}. Its mass and radius are about 0.97\,$M_{\rm \odot}$ and 0.88\,$R_{\rm \odot}$, respectively \citep{takeda2007}. Owing to its proximity and solar-like characteristics, it has been surveyed for planetary companions with the radial velocity technique \citep[e.g.][]{grether2006}, but no such companions have yet been reported. Furthermore, it has been studied in search of infrared excess indicating the presence of a circumstellar debris disk, but with null results \citep[e.g.][]{kospal2009}.

We first detected GJ~758~B with the HiCIAO high-contrast imaging
instrument \citep{hodapp2008} in angular differential imaging mode
on Subaru telescope on May 3, 2009, 
with a field of view of 20$^{\prime\prime}$ and a pixel scale of 
9.5\,mas. The images were taken in pupil-stabilized
mode in the near infrared (H~band, 1.6\,$\mu$m),
where substellar objects are expected to be bright with thermal
radiation \citep{baraffe2003,burrows2006}.
The star is known to have nine field objects \citep{carson2009}.
Data reduction of the 10 exposures of 15\,s revealed a
hitherto unknown tenth object in close separation with $5\,\sigma$ 
confidence.  A follow-up observation of 46 exposures of 9.7\,s
was obtained at the next commissioning run on August 6, 2009, in 
which the object was successfully rediscovered with $8.5\,\sigma$
confidence.  The time between epochs was long enough
to confirm common proper motion with GJ~758 and thus to establish
a gravitationally bound system.  Furthermore,
the higher quality of the second observation revealed another faint
signal of $5.6\,\sigma$ at an even smaller separation, whose nature
remains unknown until its physical existence and proper motions can
be proven with a follow-up detection at a later epoch.  
Both runs had excellent weather conditions
(0.5$^{\prime\prime}$ natural seeing in H band).

We used a locally optimized combination of images algorithm 
\citep[LOCI,][]{lafreniere2007} to maximize the efficiency of the 
angular differential imaging technique \citep[ADI,][]{marois2006}.
The signal-to-noise maps derived from the final images are 
shown in Fig.~\ref{f:LOCI}.


\section{Results}

\subsection{Proper motion analysis}

\begin{figure}[t]
\centering
\includegraphics[width=0.8\linewidth]{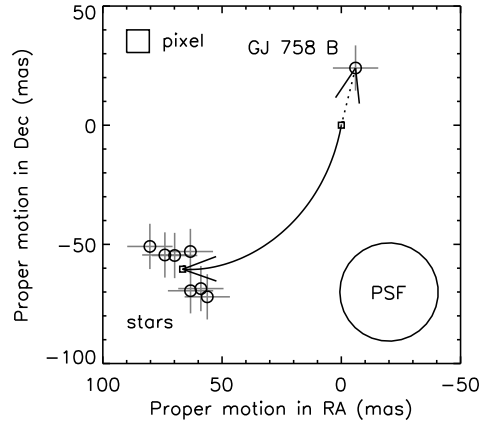}
\caption{Proper motion analysis of GJ~758's field objects.  The 
  circles show the change in the positions relative
  to GJ~758 in the May and August images for all 8 objects
  visible in both images. One-pixel error bars are drawn.
  The curved 
  arrow designates
  the predicted parallactic and proper motion for a background
  star of GJ~758 between those epochs.  GJ~758~B
  demonstrates common proper motion with its host star as well
  as some orbital motion (dotted 
  arrow), whereas the other objects conform with the behavior
  expected from
  background stars.  The size of the PSF resolution
  element and the detector pixels is provided for reference.
}
\label{f:PM}
\end{figure}

Located only 15.5\,pc away from the Sun, GJ~758~B exhibits strong
parallactic and proper motion, providing a powerful tool to distinguish
physical companions to the
target star from unrelated background stars.  In a time series of
images centered on GJ~758, a companion should remain bound to its host
while the background stars drift along the predicted background 
trajectory in unison.

Since the detections of GJ~758~B are only separated by three months,
the parallactic and proper motions are of order 0.1$^{\prime\prime}$
($\sim$10 HiCIAO pixels).  The 7 background stars visible at both
epochs were used to fine-tune the distortion correction of the
images.  As illustrated in Figure~\ref{f:PM},
this yields a tight cluster of proper motion vectors for the
background stars,
with a standard deviation on the order of the pixel scale.  GJ~758~B,
on the other hand, is found to share common proper motion with its
host star, confirming that they form a gravitationally bound system.
The companion's observed motion deviates from the proper motion
distribution of the background stars by 10$\,\sigma$, thus a chance
alignment can be excluded.

The precision of the distortion correction is 0.6 pixels in the 
central region, of the same order as the estimated precision of the
centroiding of 5$\,\sigma$ sources.  We therefore adopt a 1 pixel = 
9.5\,mas error bar for our astrometric measurements of GJ~758~B.  
The tightly constrained spread of 
the 7 background stars, which suffer from greater residual distortion
(1.3 pixels over the whole detector) and their own proper motion
(typical value 0.5 pixels in 3 months), proves this a valid
assumption.

GJ~758~B is measured to move by ($-$7.6$\pm$9.5, 24.0$\pm$9.5) 
mas along the (RA, Dec) axes relative to GJ~758 between the two
epochs.  This represents the orbital motion of the companion around
its host.  No proper motion test is available for the possible second
companion since it is detected only in one epoch.  

\begin{table}
\centering
\caption{Observed quantities of GJ~758~B and ``GJ~758~C''.} 
\begin{tabular}{lcccc}
\\
 & \multicolumn{2}{c}{GJ~758~B} & \multicolumn{2}{c}{``GJ~758~C''}\\
\hline
\hline
\multicolumn{3}{@{}l}{Photometry on August 6, 2009:} &&\\
\quad App.\ H mag.\  & \multicolumn{2}{c}{19.26 $\pm$ 0.16}
  & \multicolumn{2}{c}{18.47 $\pm$ 0.24} \\
\quad Abs.\ H mag.\ & \multicolumn{2}{c}{18.30 $\pm$ 0.16}
  & \multicolumn{2}{c}{17.51 $\pm$ 0.24} \\
\quad Contrast ($\Delta$mag)\ & \multicolumn{2}{c}{14.51 $\pm$ 0.16}
  & \multicolumn{2}{c}{13.72 $\pm$ 0.24} \\
\quad Contrast (10$^{-6}$) & \multicolumn{2}{c}{1.57 $\pm$ 0.18}
  & \multicolumn{2}{c}{3.24 $\pm$ 0.65} \\
\multicolumn{3}{@{}l}{Astrometry on May 3, 2009:} &&\\
\quad Proj.\ sep. ($^{\prime\prime}$) & \multicolumn{2}{c}{1.879 $\pm$ 0.005}
  & \multicolumn{2}{c}{---} \\
\quad Proj.\ sep. (AU) & \multicolumn{2}{c}{29.12 $\pm$ 0.08}
  & \multicolumn{2}{c}{---} \\
\quad Position angle ($^\circ$) & \multicolumn{2}{c}{197.77 $\pm$ 0.15}
  & \multicolumn{2}{c}{---} \\
\multicolumn{3}{@{}l}{Astrometry on August 6, 2009:} &&\\
\quad Proj.\ sep. ($^{\prime\prime}$) & \multicolumn{2}{c}{1.858 $\pm$ 0.005}
  & \multicolumn{2}{c}{1.188 $\pm$ 0.005} \\
\quad Proj.\ sep. (AU) & \multicolumn{2}{c}{28.80 $\pm$ 0.08}
  & \multicolumn{2}{c}{18.42 $\pm$ 0.08} \\
\quad Position angle ($^\circ$) & \multicolumn{2}{c}{198.18 $\pm$ 0.15}
  & \multicolumn{2}{c}{219.16 $\pm$ 0.08} \\
\hline
Assumed age & mass & temp. & mass & temp. \\
(Gyr) & ($M_\mathrm{Jup}$) & (K) & ($M_\mathrm{Jup}$) & (K) \\
\hline
0.7 & 10.3 & 549 & 11.7 & 631 \\
2.0 & 16.6 & 592 & 20.4 & 679 \\
4.5 & 28.6 & 623 & 35.0 & 715 \\
6.2 & 34.3 & 624 & 41.0 & 717 \\
8.7 & 39.6 & 637 & 46.5 & 733 \\
\hline
\multicolumn{5}{@{}p{0.98\linewidth}@{}}
	{\textbf{Notes.} The conversions from 
	flux to mass and effective temperature are based on the \texttt{COND}
	models by \citet{baraffe2003}.}
\end{tabular}
\label{t:photo}
\end{table}

\subsection{System age}
\label{s:age}

The age of the star is an important quantity to derive characteristics for the companion, including the mass and temperature. However, since GJ~758 is a field main-sequence solar-like star with no known connection to any co-moving stellar association, the age is very difficult to constrain. The isochronal analysis in \citet{takeda2007} places the object at an age of 0.7\,Gyr, with an upper limit of 3.8\,Gyr at 68\% confidence. The super-solar metallicity of GJ~758 ([Fe/H]$=0.14$ \citep{holmberg2009} or [Fe/H]$=0.22$ \citep{kospal2009}) might support such a relatively young age compared to the Sun, but may also just reflect a locally metal-rich natal environment. However, the activity-rotation-age calibration in \citet{mamajek2008} yields an age of 6.2\,Gyr given the rotational period of
39.0 days \citep{wright2004} and chromospheric activity $\log R^{\prime}_{\rm HK}
= -5.06$ \citep{duncan1991}.  The higher activity $\log R^{\prime}_{\rm HK} = -4.94$ given in \citet{wright2004} even results in 8.7\,Gyr.  The difference is due to monitoring during different parts of the stellar activity cycle.  The discrepancy with respect to the isochronal dating could possibly be explained with long-term activity cycles \citep[e.g.][]{janson2008} for what concerns chromospheric activity, but the rotation is harder to explain in that context.
The age from rotation alone, according to the calibration in \citet{barnes2007}, is 5.5\,Gyr.

Hence, the different age indicators for GJ~758 are discrepant. Taken together, for the purpose of our analysis, we conservatively chose the rotation-activity age estimate with the higher activity measurement of 6.2\,Gyr as the baseline case for physical interpretation. For the error bars, we considered the full non-overlapping range of 0.7 to 8.7\,Gyr.

\subsection{Physical properties of GJ~758 B}

The photometry of GJ~758~B and the candidate second companion
are listed in Table~\ref{t:photo}.  These values are based on the 
August 2009 data, since GJ~758~B is blended with a positive radial
background structure in the May 2009 data (see Fig.~\ref{f:LOCI}).
The mass and
effective temperature are derived on the basis of the \texttt{COND} 
models by \citet{baraffe2003} commonly used in
publications on substellar companions \citep{marois2008,chauvin2005}.
A sample of system ages is assumed, covering the range of
0.7--8.7\,Gyr discussed in the target properties section.

Flux measurements were performed in flat circular apertures of 5
pixels in diameter.  The loss of flux due to angular
smearing in the pupil-stabilized exposures or partial subtraction in
the ADI data reduction was carefully assessed and compensated.

For the baseline age of 6.2\,Gyr, GJ~758~B is found to have a mass of
34.3 Jupiter masses ($M_\mathrm{Jup}$) and an effective temperature of
624\,K.  At the age of 0.7\,Gyr, the mass would drop to 
10.3\,$M_\mathrm{Jup}$, placing the companion in the planetary regime,
whereas at 8.7\,Gyr it would rise to
39.6\,$M_\mathrm{Jup}$, making the companion a low-mass brown dwarf.
The effective temperature, on the other hand, is only weakly dependent
on the system age, ranging from 549\,K to 637\,K.  Given this effective
temperature, we expect GJ~758~B to have a T9 type spectral profile
\citep{leggett2009}.  This makes GJ~758~B
the coolest companion to a sun-like star ever imaged (cf.\ 750 $\pm$
50\,K for GJ~570~D orbiting a K4 triple system at 1500\,AU
\citep{burgasser2000}; 570 $\pm$ 
25\,K for Wolf 940~B orbiting an M4 star \citep{burningham2009}; 
810 $\pm$ 50\,K for HD 3651~B orbiting a more sun-like K0 star 
\citep{luhman2007}).  
The temperature range overlaps with
those of the latest-type field dwarfs
\citep{delorme2008,burningham2008,warren2007}, making GJ~758~B a
candidate for the coldest body outside the Solar System ever thermally
imaged. If the second
detected object is assumed to be a real companion, its mass estimates
go from 11.7\,$M_\mathrm{Jup}$ (also in the planetary regime) to 
46.5\,$M_\mathrm{Jup}$ and its effective temperature from 631\,K to
733\,K.

All of these temperatures are low enough to allow a strong presence of
methane in the atmosphere, which can be confirmed
with narrow-band
differential imaging in the near-infrared methane absorption bands.
The current data set, taken entirely in the broadband H filter,
cannot by itself exclude the possibility of a white dwarf companion in 
place of a brown dwarf or planet.  However, the measured contrast and
the maximum expected age of 8.7\,Gyr render this highly unlikely.  In
order for the hypothetical progenitor star to live through its 
main-sequence lifetime, evolve into a white dwarf, and cool down to the
observed H-band luminosity within 8.7\,Gyr, both the lifetime of the
star and the cooling timescale would have to be on the short end of the
scale, implying both a high progenitor mass ($\geq$2$\,M_{\odot}$)
and a high surviving white dwarf mass ($\sim$1.2$\,M_{\odot}$), two
independently unlikely assumptions \citep{fontaine2001,holberg2006}.


\section{Discussion}

Although the relative error on the measured orbital motion of GJ~758~B
is large, the data suffices to gauge the parameter space of possible
orbits and derive a ``best guess'' set of orbital parameters.  A custom
Monte Carlo simulation is used for this purpose.  The code generates a
large number (10$^6$) of orbital trajectories with
random values for eccentricity $e$, inclination $i$, argument of 
periastron $\omega$, longitude of the ascending node $\Omega$, and
a provisional semimajor axis.  The distributions are presumed
to be flat, except for the inclination, where larger angles are favored
proportionately to $\sin i$ in order to represent their higher geometric
likelihood.  The ellipse is then scaled such as to match the midpoint
between the two known companion positions within the error; this sets
the final value of the semimajor axis $a$.  Next, the physical 
velocity vector at the 
midpoint is calculated for each orbit, projected into the image plane,
and compared to the measured orbital motion.  If it fits within the 
error bars, the orbital solution is considered valid, which holds for
about 6\% of all orbits. Finally, the ensemble of valid
orbits is weighted
according to the ratio of mean and current orbital velocity, 
$\langle v\rangle/v_\mathrm{obs}$, to represent the statistical 
likelihood of finding the companion in its current position.

\begin{figure}[t]
\centering
\includegraphics[width=0.9\linewidth]{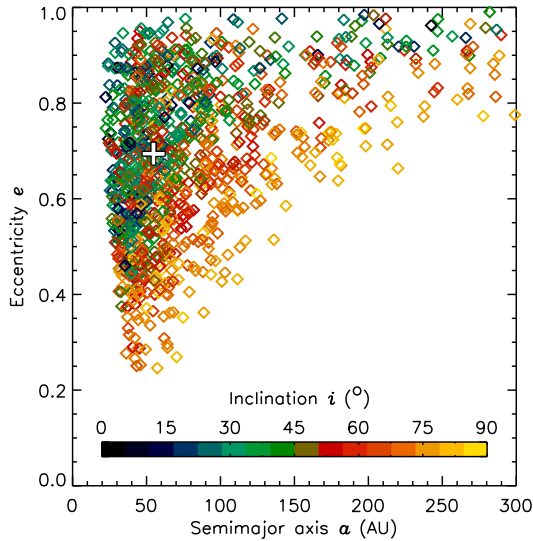}
\caption{Statistical evaluation of the orbital 
  solutions matching the observations within error bars
  among $10^6$ orbits generated by a Monte Carlo
  simulation.  The eccentricity $e$ is plotted against the 
  semimajor axis $a$ for a selection of 1000 orbits biased according to
  statistical weight.  The weight is defined as the mean orbital
  velocity divided by the velocity at the observed position, 
  $\langle v\rangle/v_\mathrm{obs}$.  The orbits are color-coded to 
  show their inclination.  The weighted median value for $(a,e)$ is
  marked with a white plus sign.
}
\label{f:orb_prop}
\end{figure}

The results are presented in Table~\ref{t:orb_prop} and 
Figure~\ref{f:orb_prop}.  Most
notably, the estimated eccentricity is very high ($e_\mathrm{WM} = 
0.691$) and constrained.  No orbital solutions below $e<0.25$
match the observations within the error bars.  The range of semimajor 
axes (33.9--118.0\,AU) places the companion in the brown dwarf desert. 

The orbital separation is too large to be the result of \emph{in
situ} formation by core accretion \citep{mordasini2008,dodson2009}.
Likewise, while gravitational instability \citep{boss2003} may in 
principle occur at these ranges \citep[in prep.]{klahr}, simulations
by \citet{stamatellos2009} suggest that such objects either rapidly
grow into full-fledged stars or are scattered to larger separations.
This leaves the possibility of formation by core accretion and 
subsequent outward transport by either gravitational scattering
\citep{veras2009} or outward migration as part of a resonant pair
of companions sharing a common disk gap \citep{crida2009}. 
 
Scattering typically leaves companions in eccentric orbits, which
matches the observation.  On the other hand, should GJ~758~C turn
out to be a real companion orbiting inwards from GJ~758~B, its
separation would be too large to be the scattering partner.  
Conversely, it would
fit the pair-wise migration scenario particularly well, especially
since it would be more massive than GJ~758~B, as the mechanism 
requires.  At any rate, such a configuration would likely be unstable
on long timescales, favoring a young system with planet-mass
companions.

\begin{table}
\centering
\caption{Estimated orbital parameters of GJ~758~B.}
\begin{tabular}{lccc}
\\
& Weighted median & 68\% likelihood\\
\hline
\hline
Semimajor axis $a$ (AU) & 54.5 & 33.9--118.0\\
Eccentricity $e$ & 0.691 & 0.497--0.866\\
Inclination $i$ ($^\circ$) & 46.5 & 24.0--67.4\\
Period $P$ (yr) & 291 & 170--658\\
\hline
\end{tabular}
\label{t:orb_prop}
\end{table}

Finally, GJ~758~B could already have formed during the star formation
phase, as a result of fragmentation of the molecular cloud \citep[e.g.]
[]{bate2002}, making it a ``failed star''.  However, such brown dwarfs
are generally ejected from the stellar system due to interactions with 
other, more massive members of the forming cluster, or undergo further
accretion to become a full-fledged star.  The resulting scarcity of 
brown dwarfs in orbit around stars \citep[conservative upper limit of 
4\% in][]{lafreniere2007b} is known as the brown dwarf desert.  
Accordingly, the hydrodynamic simulations of \citet{bate2005} favor the
formation of close binaries ($<$10\,AU) roughly equal-mass components; 
the few wide binaries with mass ratios $<$0.25 are formed by chance 
combination of two stars ejected from the cluster at the same time.  
Since a 0.97\,$M_\sun$ star like GJ~758 belongs to the most massive 
members of its birth cluster, it is unlikely to undergo ejection and 
therefore does not fit this scenario.


\section{Conclusions}

We present the discovery of a substellar companion to the star GJ~758
at a separation of 1.9$^{\prime\prime}$ by angular differential imaging
in H-band with Subaru/HiCIAO.  Common proper motion with
its host is demonstrated.  We derive ranges of 10.3--39.6
$M_\mathrm{Jup}$ for the mass and 549--637\,K for the temperature of
the companion, classifying it as either a methane-bearing high-mass 
planet or low-mass brown dwarf.
A candidate second companion is detected at a separation of 
1.2$^{\prime\prime}$, which could represent a migration
partner of GJ~758~B.

This
unique configuration of the GJ~758 system, its short dynamical timescale
(observable motion in only 3 months), and its demonstrated accessibility
by angular differential imaging, make it an excellent showcase target
for the direct investigation of the formation and evolution mechanisms
of planets and brown dwarfs.


\acknowledgements
We thank David Lafreni\`ere for generously providing us with the
source code for his LOCI algorithm, and the Subaru AO188 commissioning
team for enabling and supporting our ADI-mode observations.
This work is partly supported
by a Grant-in-Aid for Science Research in a Priority Area from MEXT,
Japan, and the U.S.\ National Science Foundation under Award No.\
AST-0901967.  This publication makes use of the SIMBAD and NStED
databases.

{\it Facilities:} \facility{Subaru (HiCIAO, AO188)}

\clearpage

\end{document}